\documentclass{PoS}

\title{The $\bar q$-$q$ potential from Bethe-Salpeter amplitudes on lattice}

\ShortTitle{The $\bar q$-$q$ potential from Bethe-Salpeter amplitudes on lattice}

\author{\speaker{Yoichi Ikeda}\\
        RIKEN Nishina Center, Wako, Saitama 351-0198, Japan\\
        E-mail: \email{yikeda@riken.jp}}

\author{Hideaki Iida\\
        RIKEN Nishina Center, Wako, Saitama 351-0198, Japan\\
        E-mail: \email{hiida@riken.jp}}

\abstract{
Potentials of quark--anti-quark pairs are studied from 
the $\bar q$-$q$ Nambu-Bethe-Salpeter (NBS) 
wave functions in quenched lattice QCD.
With the use of a method which has been recently developed 
in the derivation of nuclear forces from QCD,
we derive the $\bar q$-$q$ potentials 
with finite quark masses from the NBS wave functions.
We calculate the $\bar q$-$q$ NBS wave functions 
in pseudo-scalar and vector channels for several quark masses.
The derived potentials in both channels 
reveal linear confinement plus Coulomb potentials. 
We also discuss the quark-mass and channel dependence of 
the $\bar q$-$q$ potentials. 
}

\FullConference{The XXVIII International Symposium on Lattice Field Theory, Lattice2010\\
		June 14-19, 2010\\
		Villasimius, Italy}

\begin{document}

\section{Introduction}

An inter-quark potential is the fundamental interaction
in strongly interacting quark-gluon systems, 
which is governed by the complex dynamics of Quantum Chromodynamics (QCD). 
It is still difficult to analyze such low-energy phenomena of QCD in analytic ways,  
because the coupling constant becomes large 
at low energies and 
therefore the perturbation theory is not applicable. 
Experimentally, the linear behavior of inter-quark potentials is 
suggested by Regge slope \cite{Chiu1972a}, 
which shows the relation, $J\propto M^2$ with the spin $J$ and 
the mass $M$ of hadrons.
A naive estimation for the relation between $M$ and $J$ is 
$J= M^2/4\sigma$ with the string tension $\sigma$, 
and the value of $\sigma$ is 
about 1.3 GeV/fm from the Regge slope. 
The Coulomb force of inter-quark potentials is suggested by the analogy between
quarkonium and positronium.
In fact, Coulomb plus linear confinement behavior of inter-quark potentials 
reproduces the quarkonium spectrum well in quark models.
However, until now, there is no regolous proof of the emergence of
the linear confinement potential.

Lattice QCD simulation is the powerful tool for a numerical
investigation with strong coupling regime of strong interaction.
The 
inter-quark potential is 
one of subjects which is most actively studied on lattices.
From the analyses of Wilson loops, 
the potential for an infinitely heavy quark 
and  an anti-quark ($\bar Q$-$Q$ potential) 
can be obtained.
The $\bar Q$-$Q$ potential from lattice QCD simulations reveals 
the form of $V(r)=\sigma r - A/r$
with $\sigma=0.89{\rm GeV/fm}$ and $A=0.26$ \cite{Bali2001},
and one can take into account corrections coming from finite quark masses 
order by order with the use of the heavy quark effective field theory
such as potential nonrelativistic QCD (pNRQCD)
~\cite{Bali2001,Brown1979a,Eichten1979,Koma2007a}.

We study potentials between light quarks
and anti-quarks ($\bar q$-$q$ potentials) 
in pseudo-scalar and vector channels 
from lattice QCD simulations.
In order to explore the $\bar q$-$q$ potentials,
we apply the systematic method 
which utilize the equal-time Nambu-Bethe-Salpeter (NBS) amplitudes
to extract hadronic potentials
~\cite{Ishii2007a,Aoki2010,Nemura2008,Nemura2009,Inoue2010,Murano2009,Ikeda2010,Kawanai:10}  
to systems containing relatively light quarks and anti-quarks.
Due to the absence of the asymptotic fields of quarks, 
the reduction formula cannot be applied directly. 
Therefore, we assume that the equal-time NBS amplitudes for the $\bar q$-$q$ systems
satisfy the Bethe-Salpeter (BS) equation with constant quark masses
which could be considered as the constituent quark masses.
By using the derivation of the relativistic three-dimensional formalism from
the BS equation developed by L\'{e}vy, Klein and Macke (LKM formalism)
~\cite{Levy1952,Klein1953},
we shall obtain the $\bar q$-$q$ potentials 
without the expansion in terms of quark masses.

The paper is organized as follows.
In Sec 2, we briefly present our method to extract the $\bar q$-$q$ potentials, 
together with the lattice QCD setup.
We then show our results in Sec 3.
The $\bar q$-$q$ potentials are discussed and summarized in Sec 4.

\section{Method and lattice QCD setup}

Following the basic formulation to extract
the nuclear force~\cite{Ishii2007a,Aoki2010},
we briefly show how to extract the $\bar q$-$q$ potentials 
on the lattice below.
We start with the effective Schr\"{o}dinger equation for 
the equal-time Nambu-Bethe-Salpeter (NBS) wave function $\phi(\vec{r})$:
\begin{equation}
-\frac{\nabla^2}{2\mu} \phi(\vec{r}) + \int d\vec{r}' 
U(\vec{r},\vec{r'}) \phi(\vec{r'})
= E \phi(\vec{r}),
\label{eff-Sch-eq}
\end{equation}
where $\mu(=m_q /2)$ and $E$ denote the reduced mass of the $\bar q$-$q$ system
and the non-relativistic energy, respectively.
For the two-nucleon case, it is proved that the effective Schr\"{o}dinger
equation is derived by using the reduction formula~\cite{Aoki2010}. 
Due to the absence of the asymptotic fields for confined quarks, 
we suppose that
the $\bar q$-$q$ systems satisfy the BS equation with their 
constant quark masses. 
In this study, constant quark masses $m_q$ are determined by 
half of vector meson masses $m_V$, i.e., $m_q=m_V/2$.  
Then, one finds 
Eq. (\ref{eff-Sch-eq}) by applying LKM method to the BS equation.

The non-local potential $U(\vec{r},\vec{r'})$ can be expanded
in powers of the relative velocity $\vec{v}=\nabla/\mu$ of $\bar q$-$q$ systems
at low energies,
\begin{eqnarray}
U(\vec{r},\vec{r'}) & = &
V(\vec{r}, \vec{v}) \delta(\vec{r} - \vec{r'})  \nonumber \\
& = & (V_{LO}(\vec{r})+V_{NLO}(\vec{r})+ \cdots) \delta(\vec{r}-\vec{r'}),
\end{eqnarray}
where the $N^n LO$ term is of order $O(\vec{v}^n)$.
At the leading order, one finds
\begin{equation}
V(\vec{r}) \simeq V_{LO}(\vec{r}) = \frac{1}{2\mu}\frac{\nabla^2 \phi(\vec r)}{\phi(\vec{r})}+E.
\label{pot}
\end{equation}

In order to obtain the NBS wave functions of the $\bar q$-$q$ systems 
on the lattice,
let us consider the following equal-time NBS amplitudes 
\begin{eqnarray}
& & \chi (\vec x + \vec r, \vec x, t-t_0; J^{\pi}) =
\left\langle 0 \right|
\bar{q}(\vec x + \vec r, t) \Gamma q(\vec x, t)
\overline{{\cal J}}_{\bar qq}(t_0; J^{\pi})
\left| 0 \right\rangle \nonumber \\
& & \ \ \ \ \ \ \ \  = 
\sum_{n}A_n \left\langle 0 \right|
\bar{q}(\vec x+\vec r, t) \Gamma q(\vec x, t)
\left| n \right\rangle 
\ e^{-M_n(t-t_0)},
\label{4-point}
\end{eqnarray}
with the matrix elements
\begin{equation}
A_n = \left\langle n \right| 
\overline{{\cal J}}_{\bar qq}(t_0; J^{\pi})
\left| 0 \right\rangle.
\end{equation}
Here $\Gamma$ represents the Dirac $\gamma$-matrices,
and $\overline{{\cal J}}_{\bar qq}(t_0; J^{\pi})$ denotes a source term
which creates the $\bar q$-$q$ systems 
with spin-parity $J^{\pi}$ on the lattice.
The NBS amplitudes in Eq. (\ref{4-point})
is dominated by the lowest mass state of mesons with the mass $M_0$
at large time separation ($t \gg t_0 $):
\begin{eqnarray}
\chi(\vec r, t-t_0; J^{\pi}) 
&=&
\frac{1}{V}
\sum_{\vec x}
\chi (\vec x+\vec r, \vec x, t-t_0; J^{\pi}) \nonumber \\
&\rightarrow&
A_0 \phi (\vec r; J^{\pi}) e^{-M_0(t-t_0)},
\label{4-point2}
\end{eqnarray}
with $V$ being the box volume.
Thus, the $\bar q$-$q$ NBS wave function is defined by the spatial correlation of
 the NBS amplitudes.

The NBS wave functions in S-wave states are obtained under 
the projection onto zero angular momentum ($P^{(l=0)}$),
\begin{equation}
\phi(\vec r; J^\pi) =
\frac{1}{24}\sum_{g \in O} P^{(l=0)} 
\phi (g^{-1} \vec r; J^\pi),
\label{BS-wave}
\end{equation}
where $g \in O$ represent 24 elements of the cubic rotational group,
and the summation is taken for all these elements.
Using Eq. (\ref{pot}) and Eq. ({\ref{BS-wave}}),
we will find the $\bar q$-$q$ potentials and 
NBS wave functions from lattice QCD.


Simulation setup is as follows. 
We employ quenched QCD with
the standard plaqutte gauge action. 
Lattice size is $32^3\times 48$ and $\beta\equiv 6/g^2 = 6.0$, 
which corresponds to the physical volume $V=(3.2{\rm fm})^3$ and 
the lattice spacing $a=0.10$fm.  
We measure the $\bar q$-$q$ NBS wave functions for four different quark masses 
with hopping parameters
$\kappa=0.1520, 0.1480, 0.1420, 0.1320$: 
the corresponding 
pseudo-scalar (PS) 
meson masses $m_{\rm PS}$ in the calculation are 0.94, 1.27, 1.77, 2.53GeV, 
and vector (V) meson masses $m_{\rm V}$=1.04, 1.35, 1.81, 2.55GeV, respectively. 
The number of configurations is 100 for each quark mass. 
For the source operator of mesons, we use wall source. 
We fix the gauge,
because $q$ and $\bar q$ operators are spatially separated at the time slice of source and sink, 
and we adopt Coulomb gauge in the calculation. 

\section{Numerical results for the $\bar q$-$q$ potentials}

\begin{figure}[t]
\includegraphics[width=0.5\textwidth,clip]{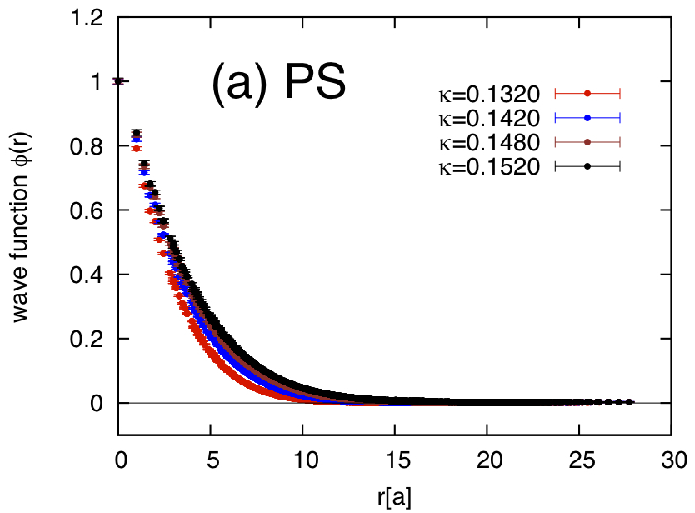}
\includegraphics[width=0.5\textwidth]{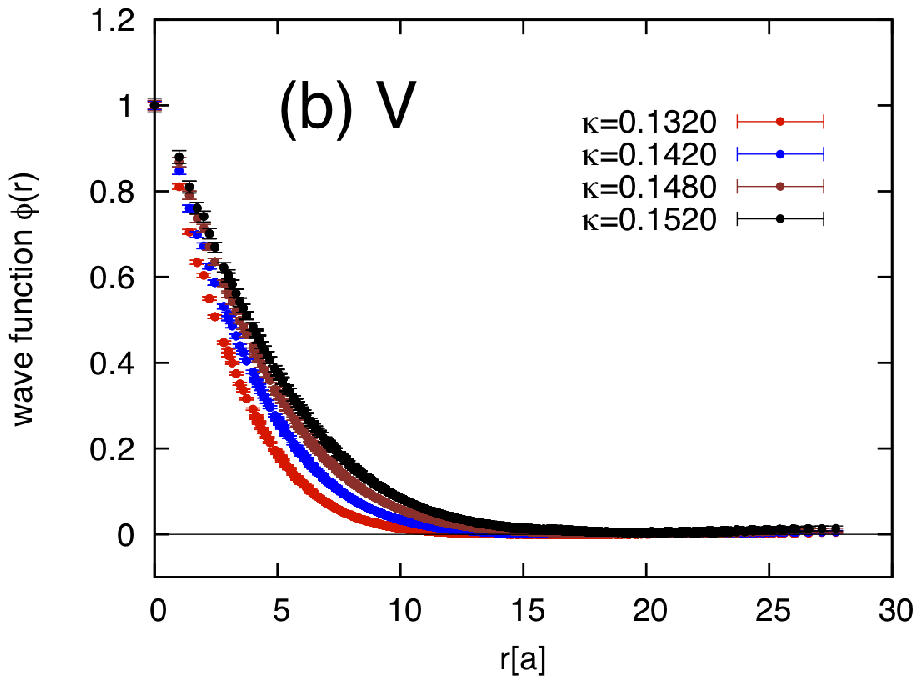}
\label{fig1}
\caption{
The $\bar q$-$q$ NBS wave functions in PS(a) and V(b) channels. 
The wave functions are normalized at origin. 
All the wave functions are localized in the box and indicate the bound states.
 }
\end{figure}
First, we show the numerical results of the NBS wave functions in Fig.~1. 
Figure 1(a) and 1(b) are the NBS wave functions 
for each quark mass in PS and V channels, respectively, 
at the time slice $t=20$.
The NBS wave functions mostly vanish
at $r=1.5$fm for all quark masses in both channels. 
This indicates that 
the spatial volume $V=(3.2{\rm fm})^3$ is enough 
for the present calculations. 
The size of a wave function with a lighter quark mass becomes smaller than 
that with a heavier one.
Comparing the results in PS and V channels, 
little channel dependence between PS and V channels is found, 
although the quark-mass dependence of the wave functions 
is a bit larger for V channel.  

In Fig.~2, we show the Laplacian parts of $\bar q$-$q$ potentials in Eq.~(2.3),
$\nabla^2 \phi(r)/\phi(r)$, 
for each quark mass and channel.
Figure 2(a)  
represents $\nabla^2 \phi(r)/\phi(r)=2\mu(V(r)-E)$  
in PS channel for each quark mass at the time slice $t=20$. 
As shown in Fig.~2, 
one can 
see that the potential form is similar to that obtained from Wilson loop, 
namely, the potential form looks like linear plus Coulomb form, 
although the derivation of the potentials 
is largely different between these two methods. 
Figure 2(b) represents  
$\nabla^2 \phi(r)/\phi(r)$ in V channel for each quark mass 
at the same time slice $t=20$. 
The basic properties are same as that in PS channel, 
although quark mass dependence is a bit larger for V channel. 

Figure 3(a) and 3(b) are the plots of the potentials with arbitrary energy shifts $E$, 
i.e, $V(r)-E$$=$$\nabla^2 \phi(r)/(2\mu \phi(r))$ 
in PS and V channels, respectively, 
for each quark mass at the time slice $t=20$. 
Note that the quark mass $m_q (= 2\mu)$ is determined 
by the half of vector meson mass, $m_q = m_V/2$, 
as mentioned in the previous section.
We fit the analytic function to 
the data in Fig.~3(a) and 3(b). 
In the present study, we choose the linear + Coulomb (+ constant) form, i.e., 
$V(r)=-A/r+\sigma r + C$, as the analytic function. 
The fitting results are listed in Table 1. 
As shown in Table 1, we find 
that the string tension $\sigma$ moderately increases as 
quark mass increases in both channels. 
Quark mass dependence of the string tension in PS channel is 
larger than that in V channel. 
The string tension at the heaviest quark mass, $m_{\rm PS}=2.53$GeV, 
is 950 (1011) MeV/fm in PS (V) channel. 
These values are roughly consistent with the value in heavy quark limit predicted from 
Wilson loop. 
On the other hand, 
Coulomb coefficient has significantly large quark-mass dependence in both channels
and is larger in PS channel than that in V channel.

\begin{figure}[t]
\label{fig2}
\includegraphics[width=0.5\textwidth]{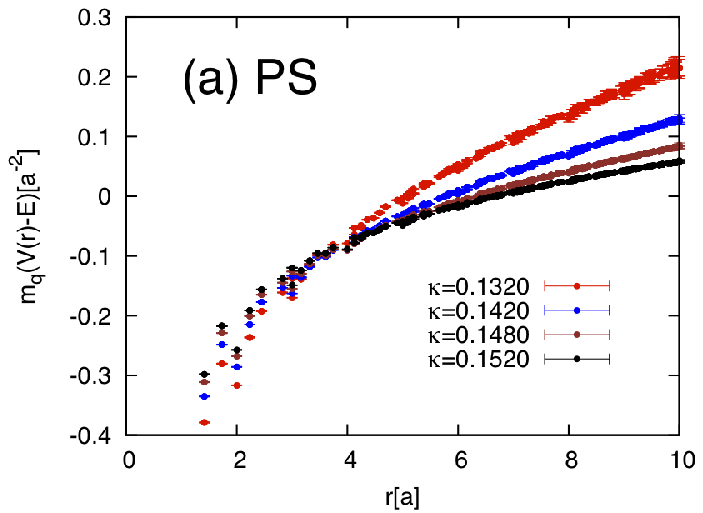}
\includegraphics[width=0.5\textwidth]{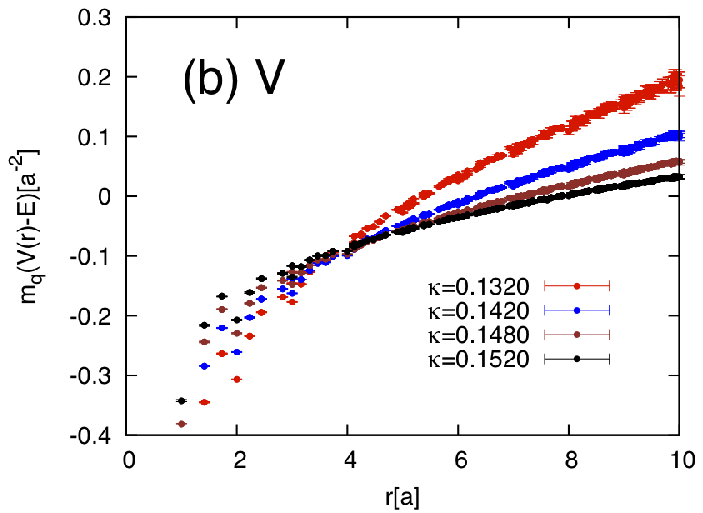}
\caption{Plots of $\nabla^2 \phi(r)/\phi(r) = 2\mu(V(r)-E)$  
in PS channel (a) and V channel (b) for 
each quark mass. 
The potential form shows Coulomb + linear behavior. }
\end{figure}
\begin{figure}[t]
\includegraphics[width=0.5\textwidth]{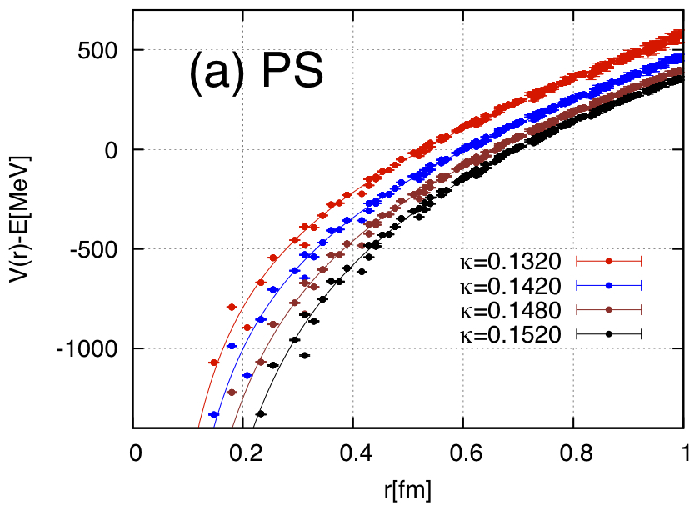}
\includegraphics[width=0.5\textwidth]{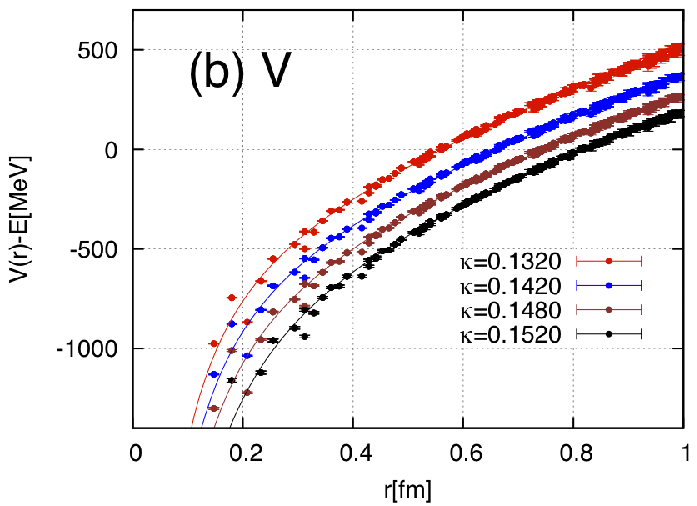}
\label{fig3}
\caption{Plots of the potential with
arbitrary constatnt energy shift $V(r)-E=\nabla^2 \phi(r)/(2\mu\phi(r))$   
in PS channel (a) and V channel (b) for 
each quark mass. 
}
\end{figure}

\begin{table}
\begin{center}
\begin{tabular}{|c|c|c|c|c|c|}
\hline
\multicolumn{3}{|c|}{Pseudo-scalar} & \multicolumn{3}{|c|}{Vector} \\
\hline
$m_{\rm PS}$ (GeV) & $\sigma$ (MeV/fm) & $A$ (MeV$\cdot$ fm) &$m_{\rm V}$ (GeV)& $\sigma$ (MeV/fm) & $A$ (MeV$\cdot$ fm) \\
\hline
2.53 & 950 & 155 &   2.55 & 1011 & 123 \\
1.77 & 878 & 193 &   1.81 & 951   & 136 \\
1.27 & 821 & 250 &   1.35 & 920   & 156 \\
0.94 & 762 & 329 &   1.04 & 914   & 182 \\
\hline
\end{tabular}
\caption{
The Fitting results of the potentials in Fig.~3. 
The data is fitted by the fit function $V(r)=-A/r+\sigma r +C$. 
}
\label{tab1}
\end{center}
\end{table}

\section{Discussion and summary}
We have studied the anti-quark--quark ($\bar q$-$q$) potentials 
from the $\bar q$-$q$ Nambu-Bethe-Salpeter (NBS) wave functions. 
For this purpose, we have utilized the method which has been recently developed in the 
calculation of nuclear force from QCD \cite{Ishii2007a,Aoki2010}. 
We have calculated the NBS wave functions 
for $\bar q$-$q$ systems with four different quark masses in pseudo-scalar 
and vector channels
and obtained the $\bar q$-$q$ potentials with finite quark masses
through the effective Sch\"{o}dinger equation.
As a result, we find Coulomb + linear form of the $\bar q$-$q$ potentials like 
the infinitely heavy $\bar Q$-$Q$ potential obtained from Wilson loop. 
By fitting the results, we have obtained the string tension and Coulomb coefficient, 
and found the quark mass dependence of these coefficients. 
We have found 
that the string tension moderately depends on the quark mass.
On the other hand, Coulomb coefficient decreases as quark mass increases. 
We have also checked
the volume and the cutoff dependence of the NBS wave functions and 
the $\bar q$-$q$ potentials. 
Then, we have found 
that the result shown here does not change quantitatively,
although we do not show these checks here. 
 
This is the first step 
to study the $\bar q$-$q$ potentials from the NBS wave functions,
and the main purpose of the present study  
to show that the method is applicable to 
the $\bar q$-$q$ potentials.
We find that the obtained $\bar q$-$q$ potential has 
basic property of that obtained from Wilson loop. 
Therefore, the method can be used for the study of the $\bar q$-$q$ potentials 
with finite quark masses. 
Since the efficiency of this method is confirmed, 
there are many extensions by using this method such as 
the dynamical calculations of the $\bar q$-$q$ potentials, 
the $\bar q$-$q$ potentials at finite temperature, 
the $3q$ potential with finite quark masses, 
color non-singlet $q$-$q$ potentials, and so on.
The results of these extensions inter-quark potentials will be reported elsewhere.

\section{Acknowledgment}
The aouthors 
thank S.~Aoki, T.Doi, T.~Hatsuda, T. Inoue, N. Ishii,
K. Murano, H. Nemura, K. Sasaki and S.~Sasaki
for the fruitful discussion.
Y.I. also thanks N. Kaiser, A. Laschka and W. Weise for the useful discussion.
The calculations were performed mainly by using the NEC-SX9 and SX8R at Osaka University, and 
partly by RIKEN Integrated Cluster of Clusters (RICC) facility.
This work is supported 
by the Japan Society for the Promotion of Science,
Grant-in-Aid for Scientific Research on Innovative Areas (No. 2004:
20105001, 20105003).


\providecommand{\href}[2]{#2}\begingroup\raggedright\endgroup


\end{document}